# Direct measurement of the magnetic field effects on carrier mobilities and recombination in tri-(8-hydroxyquinoline)-aluminum based light-emitting diodes

Feng Li,[1,a)] Linyuan Xin[1], Shiyong Liu[2], Bin Hu[3]

[1]State Key Lab of Supramolecular Structure and Materials, Jilin University, 2699 Qianjin Avenue, Changchun 130012, People's Republic of China

[2]State Key Lab of Integrated Optoelectronics, Jilin University, 2699 Qianjin Avenue, Changchun 130012, People's Republic of China

[3]Department of Materials Science and Engineering, University of Tennessee, Knoxville, Tennessee 37996, USA

**Abstract:** The magnetic field effects on the carrier mobilities and recombination in tri-(8-hydroxyquinoline)-aluminum ($Alq_3$) based light-emitting diodes have been measured by the method of transient electroluminescence. It is confirmed that the magnetic field has no effect on the electron and hole mobilities in $Alq_3$ layers and can decrease the electron-hole recombination coefficient. The results imply that the dominant mechanism for the magnetic field effects in $Alq_3$ based light-emitting diodes is the interconversion between singlet e-h pairs and triplet e-h pairs modulated by the magnetic field when the driving voltage is larger than the onset voltage of the electroluminescence.

[a)]Author to whom correspondence should be addressed; Electronic mail: lifeng01@jlu.edu.cn

The effects of magnetic fields (MFEs) on the emission and current in organic light emitting diodes (OLEDs) with nonmagnetic electrodes have been observed in 2003.[1] Since then, many groups have found that the MFEs are general phenomena which can be observed in a range of both low-molecular weight molecules and high-molecular weight polymers.[2-25] Despite the large progress in the research of MFEs, its underlying mechanism still deserves further investigations. Up to date, two competing basic models have been proposed, one is the excitonic model,[8-11] and the other is the bipolaron model.[12] The exciton model bases on that the magnetic field can either modulate the interconversion between singlet e-h pairs ($e\text{-}h^1$) and triplet e-h pairs ($e\text{-}h^3$) or modulate the reaction between $e\text{-}h^3$ and free carriers. The e-h pair is referred to the electron and hole that locate on different molecules and have Coulombic attraction to each other. Prigodin *et al.*[8] and Bergeson *et al.*[9] have suggested that organic magnetoresistance (OMAR) is due to the change in the e-h recombination coefficient caused by the change in the singlet-triplet mixing. Desai *et al.*[10] find that OMAR occurs only when there is light emission from the devices, which suggests that the OMAR is related to exciton formation. They and Hu *et al.* [11] suggest that the effect may be ascribed to the reaction between $e\text{-}h^3$ and free carriers. The bipolaron mode bases on the formation of doubly charged bipolarons during the hopping transport through the organic film, which was proposed by Bobbert *et al.*[12] The bipolaron model considers the inter-charge spin interaction and suggests that the mobilities are a function of inter-charge spin interaction. Therefore, it becomes crucial to experimentally test the MFEs on carrier mobilities and recombination to understand the underlying mechanism of MFEs in organic semiconductors.

In this work, we have measured the MFEs on carrier mobilities and e-h recombination process in tri-(8-hydroxyquinoline)-aluminum ($Alq_3$) based OLEDs by using transient electroluminescence (EL) method [26-29]. Compared to the method using constant voltage or current source, the transient EL method has the following merits in the MFEs experiments. First, the pulse EL signals of the OLEDs driven by the pulse voltage have little degradation, which is important to the reproducibility of the experiments. Second, the delay time ($t_d$) between the application of the pulse voltage and the intersection of the rising edge of the pulse EL signal and the baseline is related to the carrier mobilities. Third, the decay of the EL pulse after removing the pulse voltage includes the information of the e-h recombination. Forth, the flat period of the pulse EL signal can elucidate the MFEs on EL ($MFE_{EL}$). Thus, the MFEs on the carrier mobilities, e-h recombination and EL can be obtained from the transient EL measurements, simultaneously. The OLEDs used in this work were fabricated by using multiple-source organic molecular beam deposition method.[30]

To test the MFEs on the electron mobility in $Alq_3$ layer, device A with the structure of ITO/ N, N′ -di-1-naphthyl-N, N′ -diphenylbenzidine (NPB) (40 nm)/tri-(8-hydroxyquinoline)-aluminum ($Alq_3$) (60 nm)/LiF (0.5 nm)/Al (100 nm) has been made, as shown in the insert of Figure 1 (a). The NPB is the hole-transport layer while the $Alq_3$ functions as both emissive and electron-transport layer. Figure 1 (a) shows the full EL response of device A at the driving voltage from 2 V to 4 V with and without the magnetic field. As can be seen from figure 1 (b), the rising edges of the EL pulses with (red line) and without (blue line) the magnetic field overlap perfectly, thus the magnetic field has no effect on the delay times of $t_d$. In generally, the $t_d$ is thought to be the transit time for the injected electrons from the cathode to the recombination zone in $Alq_3$ layer of device A, because the hole mobility in NPB layer is at least one order of magnitude larger than the electron mobility in $Alq_3$ layer, the transit time for hole in the NPB layer can be neglected. The mobility of electron in $Alq_3$ layer can be calculated by $\mu_e=L/(t_dE)$, approximately, where $L$ is the thickness of $Alq_3$ layer and $E$ is the electric field in $Alq_3$ layer (the $E$ field in the NPB layer is neglected). Figure 3 shows the calculated mobility of electron in $Alq_3$ at the driving voltage from 2 V to 4 V, which is consistent with the results reported by S. Barth *et al.*[28] The magnetic field has no effect on the electron mobility in $Alq_3$ layer. This experimental result indicates that the effect of magnetic fields on the electron transport process is not the dominant mechanism for the MFEs in this $Alq_3$-based device when the driving voltage is larger that the onset voltage of the EL. Desai *et al.* suggest the magnetic fields may influence the carrier mobility caused by the scatter effect of the triplet on the carriers [10]. Since the $t_d$ is only related to the electron transport process in $Alq_3$ layer and is obtained before the creation of the excitons in our experiment, the scatter effect of the triplet on the mobility of electrons derived from $t_d$ can be ignored. Thus our result that the magnetic field has no effect on the electron mobility in $Alq_3$ layer has no contradiction with the suggestion of Desai *et al.*

Figure 1 (c) shows the effect of magnetic fields on the emission of device A at the driving voltage of 4 V. The effect is positive and calculated to be $MFE_{EL}= (EL(B) -$ $EL(0))/EL(0)$=0.54 %. The values of $MFE_{EL}$ at other driving voltages are shown in the insert of Figure 1 (a). Furthermore, Figure 1 (d) shows the EL pulses at turn-off period (the falling edges) of device A with and without the magnetic field on a logarithmic scale. As can be seen, the EL decay signal after turning off the voltage pulse consists of fast and slow components. The fast decay component corresponds to the radiation decay of the preexisting (singlet) excitons and the slow component should be ascribed to the bimolecular recombination process of the accumulated charges in the recombination zone.[31,32] The EL decay curves for the device with and without the magnetic field are separated from each other in the slow component, which indicates that the magnetic field actually affects the bimolecular recombination process. Assuming the bimolecular recombination in the narrow zone in $Alq_3$ layer to be the main channel for carriers decay, that is neglecting carrier leakage through the interface, the slow component of the EL decay can be fitted by:[33]

$$\frac{1}{\sqrt{\phi_{EL}(t)}} = A + St \qquad (1)$$

Where $\Phi_{EL}$ is the EL intensity. Then the bimolecular recombination coefficient $r$ can be expressed as:[33]

$$r = \frac{(S/A)^2 ed}{2j} \qquad (2)$$

Where $e$ is the electron charge, $d$ is the thickness of the device and $j$ is the current density. The parameters of $A$ and $S$ in equation (2) can be obtained by fitting the slow component of the EL decay data by equation (1), they are the function of the magnetic field. The MFEs on the bimolecular recombination coefficient ($MFE_r=(r(B)-r(0))/r(0)$) can be calculated by equation (2) and the calculated values are shown in the insert of figure 4 (a). The calculated results show that the recombination coefficient $r$ can be decreased by the magnetic field of 200 mT. In the presence of a magnetic field, the interconversion between the triplet e-h pairs and singlet e-h pairs is decreased, and the long-lived triplet e-h pairs have more likelihood of dissociation, which leads to the decrease of the total e-h recombination coefficient.[8,9] It should be noticed that the MFEs for the driving voltage from 1.6 V to 15 V have been tested and the results are similar with those showed in figure 1.

To test the MFEs on the hole mobility in $Alq_3$ layer, device B with the structure of ITO/NPB (10nm)/$Alq_3$ (80nm)/NPB (40nm)/$Alq_3$:Rubrene (20nm)/LiF (0.5nm)/Al (100nm) has been made, as shown in the insert of Figure 2 (a). The 10 nm thick NPB and 80 nm thick $Alq_3$ layers act as the hole-injecting and hole-transporting layers, respectively. The 40 nm thick NPB layer is used to block the electrons injected from the cathode. The 20 nm thick $Alq_3$ doped with 1% Rubrene layer is the light-emitting layer. The mobilities of holes in NPB and electrons in $Alq_3$ are at least one order of magnitude larger than the hole mobility in $Alq_3$. For this reason, the transit time of the electrons in the Rubrene doped $Alq_3$ layer and the transit time of the holes through both NPB layers can be neglected. The delay time ($t_d$) is approximately equal to the transit time of holes through the 80 nm thick $Alq_3$ layer. Thus the hole mobility in $Alq_3$ layer can be calculated by $\mu_h=L/(t_dE)$, approximately, where $L$ is the thickness of $Alq_3$ layer (80 nm) and $E$ is the electric field in the $Alq_3$ layer. The calculated values shown in figure 3 are of the order $10^{-6}$ $cm^2/Vs$, which is in agreement with the results of L. Lin *et al.*[34] As can be seen from figure 2 (b), the rising edges of the EL pulses with (red line) and without (blue line) the magnetic field overlap perfectly, which verifies the magnetic field has no effect on the hole mobility in $Alq_3$ layer. The EL decay curves for the device with and without the magnetic field are separated from each other in the slow component as shown in figure 2 (d). Therefore, it confirms than the magnetic field affects the bimolecular recombination process again. The similar results can be obtained for the driving voltage from 6 V to 12 V. Figure 2 (c) shows the effect of magnetic fields on the emission of device B at the driving voltage of 9 V. The effect is positive and calculated to be 0.51 %. The values of $MFE_{EL}$ at other driving voltages are shown in the insert of Figure 2 (a). Figure 4 (b) shows the fitted results of the EL slow decay component of the device B by equation (1). The values of $MFE_r$ are calculated by equation (2) and are shown in the insert of figure 4 (b). The calculated results indicate the recombination coefficient $r$ can be decreased by the magnetic field.

In summary, we have performed the transient EL experiments to directly measure the MFEs on carrier mobilities and recombination in $Alq_3$ based OLEDs. The rising edges of the EL pulses overlap perfectly while the EL decay curves (the falling edges) are separated in the slow component with and without the magnetic field. The results provide the evidences that the magnetic field has no effect on the electron and hole mobilities in $Alq_3$ layer. Therefore, it can be suggested that the effect of magnetic fields on the carrier transport process is not the dominant mechanism when the driving voltage is larger than the onset voltage of the EL, and the magnetic field actually can decrease the e-h recombination coefficient.

We are grateful for financial support from National Natural Science Foundation of China (grant numbers 60878013 and 60706016) and “111” Program (B06009).

## Figure captions

**FIG. 1.** The full transient EL response with (200 mT, red line) and without (blue line) the magnetic field of device A (a). The detail of the EL pulses at rising edges, flat and falling edges period are shown in (b), (c) and (d), respectively. The repetitive frequency and the width of the driving pulse are 1 KHz and 7 μs, respectively. Insert of (a) shows the structure of device A.

**FIG. 2.** The full transient EL response with (200 mT, red line) and without (blue line) the magnetic field of device B (a). The detail of the EL at rising edges, flat and falling edges period are shown in (b), (c) and (d), respectively. The repetitive frequency and the width of the driving pulse are 1 KHz and 7 μs, respectively. Insert of (a) shows the structure of device B.

**FIG. 3.** Electric field dependence of charge carrier mobilities in $Alq_3$ thin film.

**FIG. 4.** The slow component of EL decay curves for device A (a) and device B (b) with and without the magnetic field, plotted in $(\Phi_{EL})^{-1/2}$ vs time scale. The data are fitted (solid straight line) by Eq. (1).

Figure 1

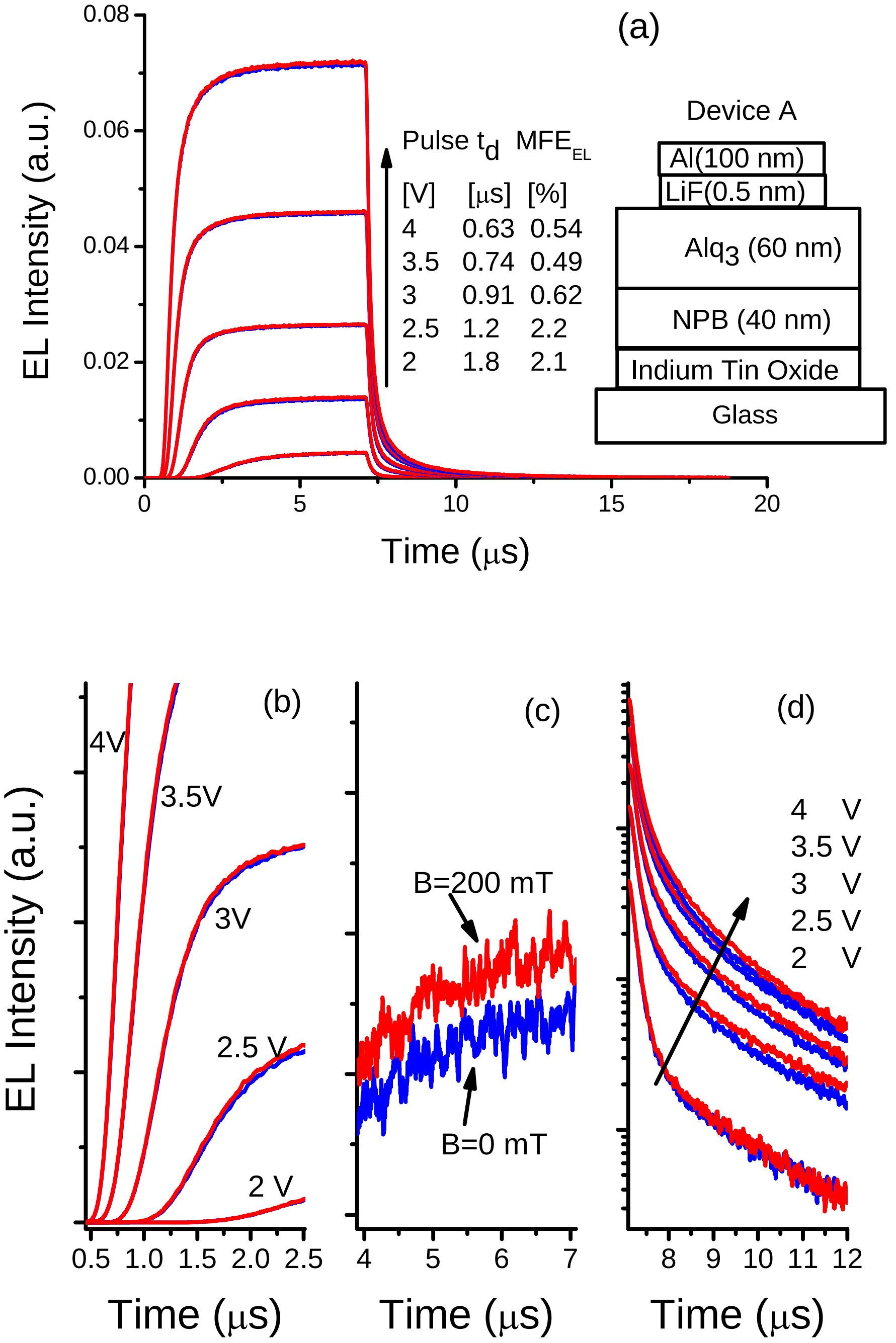

Figure 2

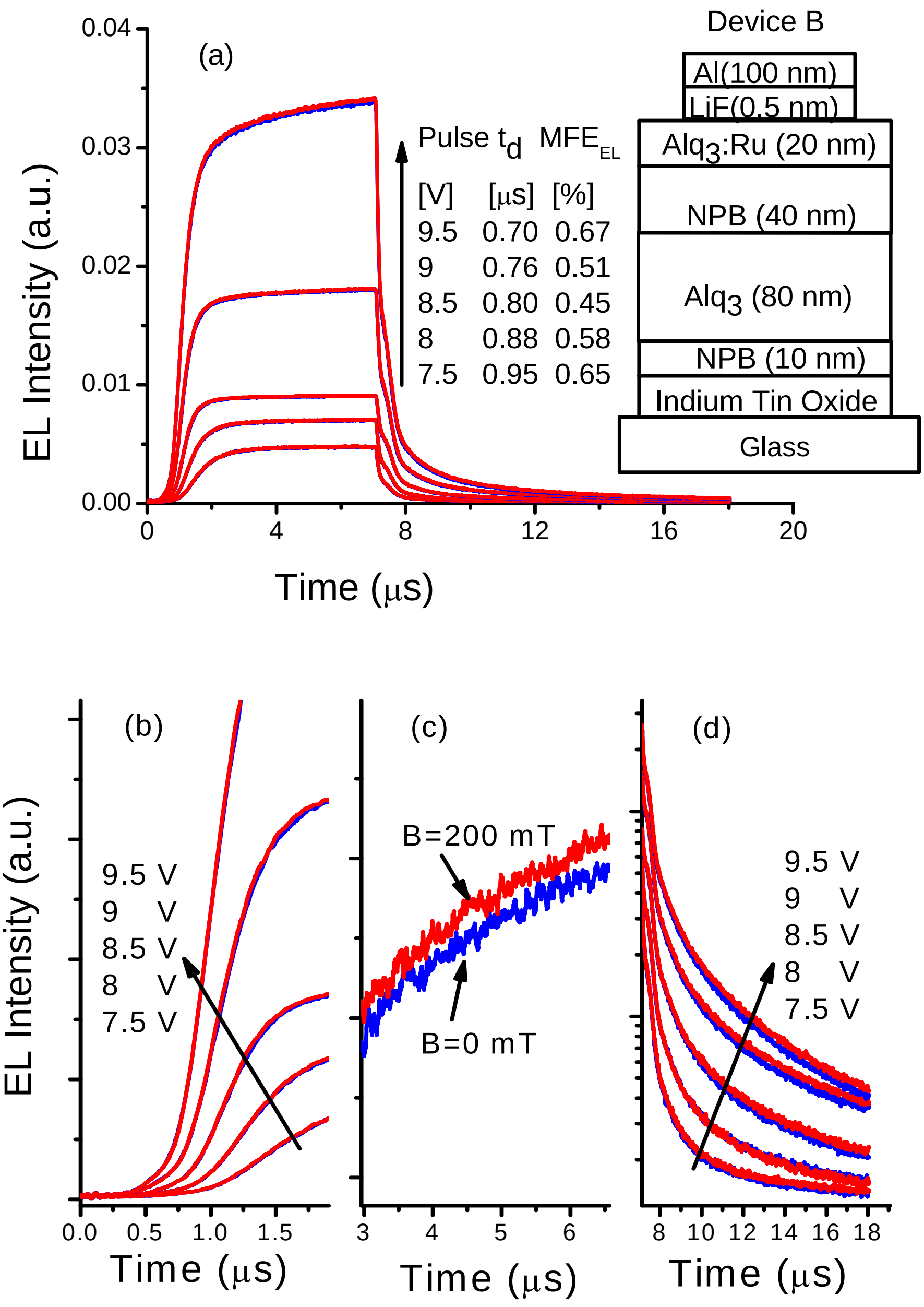

Figure 3

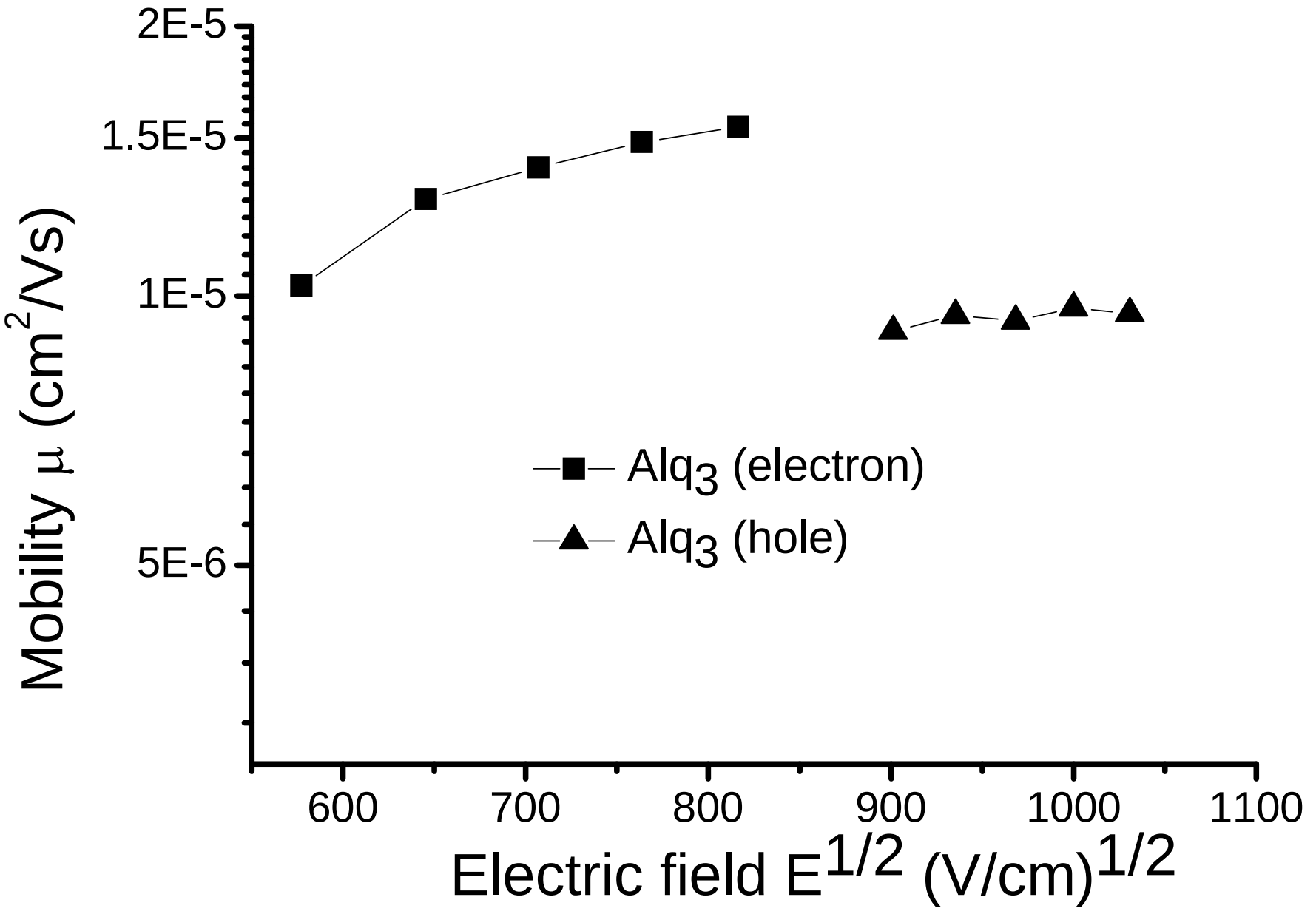

Figure 4

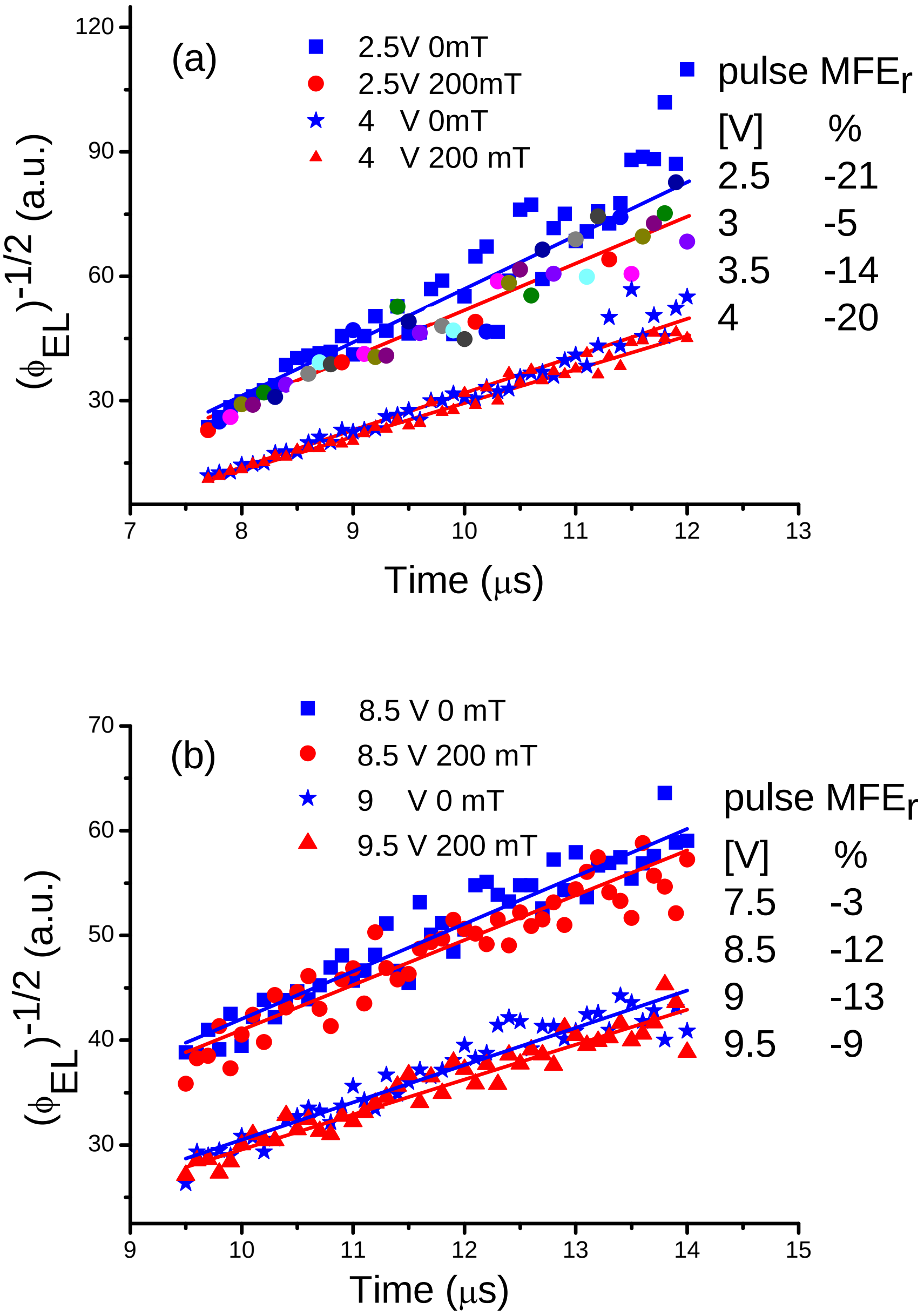